\documentclass[ aip, amsmath,amssymb, reprint, ]{revtex4-2}

\usepackage{hyperref}
\usepackage{graphicx}
\usepackage{bm}
\usepackage[utf8]{inputenc}
\usepackage[T1]{fontenc}
\usepackage{newtxtext}
\usepackage{newtxmath}
\usepackage{kantlipsum}
\usepackage{cancel}

\newcommand*\overbar[1]{%
  \vbox{%
    \hrule height 0.5pt
    \kern0.25ex
    \hbox{%
      \kern-0.05em
      \ifmmode#1\else\ensuremath{#1}\fi
      \kern-0.05em
    }
  }
}

\def\rr{{\bm r}}
\def\RR{{\bm R}}
\def\DD{{\bm D}}

\def\dd{{\bm d}}
\def\qq{{\bm q}}
\def\order{\mathcal O}
\def\v{\varv}
\def\half{{0.5}}

\begin{document}

\title{Birth and ephemeral life of pseudopotential alchemy}
\author{Stefano Baroni}
 \affiliation{SISSA -- Scuola Internazionale Superiore di Studi Avanzati, Trieste, Italy}
 \affiliation{ CNR-IOM –- Istituto Officina dei Materiali, Trieste, Italy}

 \email{baroni@sissa.it}

\date{\today}

\begin{abstract}
The theory of band offsets at semiconductor interfaces has been one of Alfonso Baldereschi’s most cherished topics, to which he has made significant contributions—--both directly and by enticing peers and disciples, including myself, to delve deeply into it. In this tribute, I recount how a brilliant idea of his for modeling band offsets in a class of semiconductor heterojunctions led to one of our most cited works and brought about the birth of \emph{pseudopotential alchemy}. This model provides a powerful tool to both compute and fathom the electronic and structural properties of composite semiconductors, such as heterojunctions and alloys, by treating them as small perturbations with respect to a \emph{virtual crystal}---a system in which different chemical elements occupying crystallographically equivalent lattice sites are represented by an average pseudopotential. This note is intended as a tribute to the fond memories I have of the time spent with, and the science learned from, Alfonso. No attempt will be made at bibliographic completeness in the niche field of pseudopotential alchemy, let alone the much broader one of semiconductor heterojunctions.
\end{abstract}

\maketitle

The rapid diffusion of the \emph{Molecular Beam Epitaxy} (MBE) technology \cite{McCray:2007} in the eighties was pivotal in enabling the widespread fabrication of semiconductor heterostructures, thus paving the way to the emergence of nanosciences in the two-thousands. This development sparked the interest of solid-state theorists, with the young Alfonso Baldereschi standing out as a towering figure, who sought to understand, predict, and eventually engineer their electronic properties. Alfonso was not just any theorist. To mark the difference between abstract theory and the field of which he was an internationally renowned leader, he liked to define the latter as the ``theory of \emph{physical phenomena}''. And physical phenomena are to be \emph{observed} either before being fathomed or after being predicted. In both cases, observation needs increasingly sophisticated apparatuses, which drove Alfonso's instrumental role in the acquisition of Italy's first MBE machinery for the TASC National Laboratory in Trieste---where he had secured a second chair at the University in 1981 and continued his influential work until his retirement in 2016, after having held a similar position in Lausanne until 2011.

Following Alfonso’s path, in 1984 I relocated to Trieste from Lausanne, where my long and profoundly shaping acquaintance with him had begun in 1979. The late eighties were creatively significant for me, marked by the development of density-functional perturbation theory (DFPT) \cite{Baroni1987a,Giannozzi1991a,Baroni2001}-—-a result of the strong and extremely fruitful collaboration between Trieste and Lausanne in those years---and my deep involvement in semiconductor physics, into which Alfonso prodded me, starting with our most cited joint work \cite{Baldereschi:1988}.

In the mid-eighties, a much-debated issue was whether band offsets at semiconductor heterojunctions are intrinsic bulk properties of the constituent materials or if they critically depend on the orientation and morphology of the interface. As of 1987, the prevailing view was mixed: while general theoretical considerations based on the long-range nature of the Coulomb interaction suggested that electrostatic potential lineups—and therefore band offsets—at semiconductor interfaces may indeed depend on the interface structure \cite{Kleinman1981}, experimental evidence \cite{Wang1985}, first-principles calculations on the paradigmatic GaAs/AlAs system \cite{Bylander1987}, and theoretical models applicable across different chemical compositions \cite{VanDeWalle:1987} seemed to suggest the opposite. The situation was intriguing: theory clearly showed that band offsets could depend on the complex nature of the interfacial structure, which in turn depends on the details of the fabrication process—-why shouldn't it?-—yet experiment and computation seemed to indicate the opposite, at least in the case of GaAs/AlAs, which was the most studied example.

\emph{The purpose of computation is insight, not numbers} \cite{Hamming1962}: if current calculations do not explain what appears to be a fortuitous coincidence, it is because the logical consequences that could be drawn from them have not yet been thoroughly worked out. Armed with this belief, Alfonso was determined to dig deep into the problem and involved me in our first joint paper on semiconductor interfaces in 1987 \cite{Baldereschi:1988}, thus luring me into the field of semiconductor physics, which kept me busy for the next ten years or so. This paper, which is still now one of the most cited for the two of us, as well as for its third author, Raffaele Resta, introduced two far reaching concepts.

The first one, which was actually due to Raffaele, and is still widely used to date in the science of surfaces and interfaces, is that of \emph{macroscopic average}. This concept provided a novel tool for analyzing the $z$-dependence of a scalar field, such as the electron density and the electrostatic potential energy, across the junction of two semi-infinite media (one of which could simply be the vacuum, as at surfaces). By averaging a field over a period centered at a specific point, macroscopic averages effectively smooth out short-wavelength fluctuations, isolating the long-range variations that are crucial for understanding electrostatic properties at interfaces. This approach allows for a clearer distinction between bulk and interface-specific features, without requiring the definition of an ideal reference interface, while offering deeper insights into the electrostatic behavior at semiconductor heterojunctions. Leveraging the accurate numerical analysis that the concept of \emph{macroscopic average} can afford, the band offset at GaAs/AlAs heterojunctions was found to be independent of interface orientation to within 0.02 eV. It was concluded that \emph{the finding of an orientation-independent macroscopic dipole suggests the idea that for GaAs/AlAs the lineup is basically a bulk effect. Such independence is in fact obtained under the assumption that each of the two bulk solids is an assembly of elementary building blocks, and that these blocks can also be rigidly assembled to form an ideal reference interface. Starting from this reference, any orientation dependence can only be due to electronic redistributions at the interface.} Alfonso then introduced the second key idea of the paper by suggesting that such elementary building blocks could be identified with the cation-centered Wigner-Seitz cells (WSC) of the two constituents. Overall charge neutrality requires that these building blocks be neutral, while crystal symmetry demands that they carry no dipole nor quadrupole moments. Elementary electrostatics then dictates that the average electrostatic potential of a semi-infinite three-dimensional array of such building blocks, with vanishing first multipoles, is independent of interface effects. In fact, the Fourier transform of the charge-density distribution of any such \emph{finite} array is:
\begin{equation}
    \tilde\phi(\qq)=\frac{4\pi}{q^2} \tilde \rho_{WSC}(\qq) \sum_{\RR} e^{-i\qq\cdot\RR},
\end{equation}
where $\tilde \rho_{WSC}$ is the Fourier transform of the charge-density distribution within the WSC, and $\{\RR\}$ their positions. The average electrostatic potential of the array then reads:
\begin{equation}
    \begin{aligned}
        \bar\Phi &= \frac{1}{N\Omega} \lim_{\qq\to 0} \tilde \phi(\qq) \\
                 &= \lim_{\qq\to 0}\frac{4\pi}{q^2\Omega} \left (\cancel{Q} -i \cancel{\qq\cdot\dd} - \frac{1}{2}\cancel{\qq\cdot\DD\cdot\qq} - \frac{1}{2}Q_2 q^2 +\cdots \right ) \\
                 & = -\frac{2\pi Q_2}{\Omega},
    \end{aligned}
\end{equation}
where $\Omega$ is the volume of the WSC, $Q$, $\dd$, and $\DD$ its total charge, dipole, and quadrupole, $Q_2=\frac{1}{3}\int \rho'(\rr)r^2d\rr$ its second spherical moment, and $N$ their number. It follows that the macroscopic average of the electrostatic potential generated by a periodic array of localized  neutral charge-density distributions bearing no dipole nor quadrupole is well defined and independent of boundary effects. In this scenario, the potential lineup at a semiconductor interface is simply the difference between the average potentials of the two semi-infinite constituents and the band offset independent of any interface effects: $\Delta\bar\Phi(\mathrm{GaAs/AlAs)}=\frac{2\pi}{\Omega}\Bigl (Q_2(\mathrm{AlAs}) -Q_2(\mathrm{GaAs})\Bigr ) $. This finding would solve the problem of the band offset at lattice-matched homo-polar semiconductor interfaces, but, if it were general, it would leave little room to engineering the interface so as to tune the band offset across it. What about the effects of strain, such as in Si/Ge or GaAs/InAs, of heterovalency, such as in Ge/GaAs, or of a combination of the two, such as in Si/GaAs? Is there any way that we can exploit these effects, if any, to tune the band offset across a semiconductor interface? In order to answer these questions, a new approach to semiconductor heterojunctions was developed, based of \emph{pseudopotential alchemy}, and later extended to solid solutions.
 
With the establishment of the \emph{atomic pseudopotential} concept \cite{Hellmann:1935,*Phillips:1958,*Heine:1970,*Hamann:1979} and the advent of computing systems powerful enough to model the electronic structure of real materials, by the mid-seventies the virtual-crystal approximation (VCA) \cite{Nordheim1931} had gained considerable traction in semiconductor physics for addressing compositional disorder in solid solutions. A solid solution is characterized by a regular crystalline structure, where (sub-) lattice sites are occupied at random by two (or more) atomic species, say $A$ and $B$. In a nutshell, the VCA assumes that all the atoms in the disordered (sub-) lattice are represented by the same average pseudopotential, $\bar \v(\rr)=\frac{1}{2}\bigl ( \v_A(\rr)+\v_B(\rr) \bigr )$, resulting in a perfect crystal where the effects of disorder are lost, but the average properties of the alloy are hopefully represented with acceptable accuracy. Of course, the VCA representation of a semiconductor heterojunction would entirely overlook any interface-specific features, including band offsets. If the VCA is sufficient to describe the average macroscopic properties of a composite semiconductor—whether an alloy or a heterostructure—it must be because the difference between $\v_A$ and $\v_B$ is small enough to be treated by low-order perturbation theory. We reckoned therefore that linear response theory (LRT) would probably give a good insight into the band offset problem \cite{baroni:1989Venice}. This simple idea turned out to be extremely fruitful, not just for computing the electronic properties of interfaces, but also for predicting their dependence on structural details, often using only simple back-of-an-envelope considerations on top of sophisticated first-principles calculations.

In the VCA, the bare pseudopotential acting on the electrons reads:
\begin{equation}
    \overbar{V}(\rr)=\sum_{\RR} \bar\v(\rr-\RR), \label{eq:VVCA}
\end{equation}
where the sum runs over the (sub-)lattice sites affected by substitutional disorder (as in alloys) or structured order (as in heterostructures). The actual alloy/interface bare (pseudo-) potential is therefore:
\begin{equation} \label{eq:Valloy}
    \begin{aligned}
        V(\rr) &= \overbar{V}(\rr)+ \Delta V(\rr) \\
        \Delta V(\rr) &= \sum_{\RR} \sigma_{\RR}~\v'(\rr-\RR) \\
        \v' &= \frac{1}{2}\bigl ( \v_A-\v_B \bigr ),
    \end{aligned}
\end{equation}
where $\sigma_{\RR}=1$ if the (sub-) lattice site at $\RR$ is occupied by an atom of type $A$ and  $\sigma_{\RR}=-1$ if it is occupied by an atom of type $B$. To linear order in $\Delta V$, the charge-density response to it reads:
\begin{equation}
    \Delta\rho(\rr)=\sum_\RR \sigma_\RR~\rho'(\rr-\RR), \label{eq:Delta_rho'}
\end{equation}
where $\rho'$ is the virtual-crystal charge-density response to $\v'$. $\rho'$ is a localized, neutral, charge-density distribution, whose dipole and quadrupole in a tetrahedral semiconductior vanish because of crystal symmetry. It can be easily demonstrated that, in the long-wavelength ($\qq\to 0$) limit, the Fourier transform of the electrostatic potential generated by the charge-density distribution of Eq. \eqref{eq:Delta_rho'} coincides with that produced by a dipole layer, $\sigma d=\frac{4\pi e^2}{\Omega}Q'_2$, where $e$ is the electron charge and $Q'_2$ the second spherical moment of $\rho'$, giving rise to an electrostatic potential lineup of same magnitude, independent of interface orientation or abruptness.

Consider now the interface between two heterovalent, though still approximately lattice-matched, semiconductors, such as Ge/GaAs. In this case, the appropriate virtual crystal is a zincblende $\mathrm{\langle Ge_\half Ga_\half \rangle\langle Ge_\half As_\half \rangle}$, whose cation has valence charge 3.5, while the anion has charge 4.5. The relevant (bare) localized perturbations that transform a virtual
cation ($C$) or anion ($A$) into a real or germanium one are:
\begin{equation} \label{eq:hetero_perturbations}
   \begin{split}
      \v'_C &= \frac{1}{2}(\v_\mathrm{Ga}-\v_\mathrm{Ge}) \\
      \v'_A &= \frac{1}{2}(\v_\mathrm{As}-\v_\mathrm{Ge}) .
   \end{split}
\end{equation}
The bare perturbations of Eq. \eqref{eq:hetero_perturbations} carry a net charge of $\pm 0.5$. Therefore, to first order in the perturbation, the total (bare plus screening) charge induced by them is $\pm\frac{1}{2\epsilon_\infty}$, where $\epsilon_\infty$  is the static electronic dielectric constant of the virtual crystal. It is to be stressed that this result is \emph{exact} within LRT and fully takes into account self-consistency and local-field effects.

According to the previous discussion, the Fourier transform of the total charge induced by one localized perturbation reads:
\begin{equation} \label{eq:Delta_rho'AC}
    \tilde\rho'_{A,C}(\qq)=\pm \frac{1}{2\epsilon_\infty} -\frac{1}{2}Q'_{A,C} q^2 + \order(q^3). 
\end{equation}
In the spirit of LRT, the corresponding potential lineup can be split into two contributions:
\begin{equation}
    \Delta\Phi=\Delta\Phi^{(0)}+\Delta\Phi^{(2)},
\end{equation}
due to the terms of order zero and two, respectively, in Eq. \eqref{eq:Delta_rho'AC}. The latter is by definition the lineup due to an array of neutral, symmetric, perturbations: as discussed earlier, $\Delta\Phi^{(0)}$ is independent of the orientation and abruptness of the interface. The former is instead formally equivalent to the lineup generated by an assembly of \emph{point charges} with absolute value $\frac{1}{2\epsilon_\infty}$ and \emph{does depend} on the atomic structure of the interface---its orientation, abruptness, geometric relaxation, and so forth. However, once the structure is determined, either experimentally of by independent theoretical calculations, $\Delta\Phi^{(0)}$ can be computed from elementary electrostatics. This allows, at least in principle, for engineering the interface structure to tune the band offset across the junction \cite{baroni:1989Venice}. The most spectacular demonstration of how interface engineering can tune the band offset at a semiconductor junction is the realization that an electrostatic potential drop can be induced across the interface between two like semiconductors (e.g., Ge) grown along a polar direction, such as (001) or (111), by the deposition an ultrathin double layer of a heterovalent material (e.g., GaAs) \cite{Peressi:1991}. The explanation is quite simple: the heterovalent double layer acts like a microscopic capacitor, with plates carrying a charge density of $\sigma= \pm\frac{1}{A_{lmn}\epsilon^{Ge}\infty}$, where $A_{lmn}$ is the area of the surface unit cell in the growth direction, $(nlm)$, and the plates are separated by the interplanar lattice distance, $d_{lmn}$. This prediction---regarding the dependence of band offsets at semiconductor heterojunctions and the creation of an offset at homojunctions via the deposition of ultrathin dipole layers---was soon confirmed by X-ray photoemission spectroscopy \cite{Biasiol:1992,Marsi:1992}. Using similar arguments, the effects of strain on the band offset could be fathomed\cite{Peressi:1993a}, leading to the development of a general LRT theory of band offsets at semiconductor heterojunctions, as reviewed, e.g., in Ref. \onlinecite{Peressi:1998}.

Shortly after one of the major issues in the physics of semiconductor heterostructures was thus resolved by \emph{pseudopotential alchemy}, it was realized that the very same concept could be leveraged to address the structural stability of semiconductor solid solutions. In the VCA, a solid solutions such as Si$_{0.5}$Ge$_{0.5}$ is described by the pseudopotential of Eq. \eqref{eq:VVCA}. Pseudopotential alchemy turns the VCA crystal into a real alloy through the perturbation of Eq. \eqref{eq:Valloy}, where the $\{\sigma_\RR \}$ are random variables distributed according to the Boltzmann distribution, $P[\sigma]\propto e^{-\beta E[\sigma]}$, and $E[\sigma]$ is the energy of a given configuration. Standard quantum mechanical (or density-functional) perturbation theory prescribes that the total energy of the alloy reads \cite{Baroni2001}:
\begin{equation}
    \begin{aligned}
        E[\sigma] &= E_{VCA}+\cancel{\int \rho_{VCA}(\rr)\Delta V(\rr) d\rr} \\ &\qquad + \frac{1}{2} \int \Delta\rho(\rr) \Delta V(\rr) d\rr + \order(\Delta V^3) \\
                  &\approx E_{VCA}+ \frac{1}{2} \sum_{\RR\RR'} \sigma_\RR\sigma_{\RR'} J(\RR-\RR'),
    \end{aligned}
\end{equation}
where the linear term vanishes because $\sum_\RR \sigma(\RR)=0$ and $J(\RR-\RR')=\int \rho'(\rr-\RR) \v'(\rr-\RR')d\rr'$. It follows that, to the lowest significant order in perturbation theory, the energy of the alloy can be mapped onto an Ising model with long-range interactions, whose coupling constants, $J(\RR-\RR')$, can be calculated using DFPT, and whose thermodynamic properties can be determined by standard Metropolis Monte Carlo simulations. The effects of macroscopic strain and microscopic atomic distortions from the VCA equilibrium configuration can be incorporated by introducing a generalized perturbation, in which the ``alchemical'' variables, $\{\sigma_\RR\}$, are coupled with the atomic displacements $\{u_\RR\}$. Solid solutions of arbitrary binary, ternary, or quaternary compositions, such as Si$_x$Ge$_{1-x}$ \cite{deGironcoli:1991}, Ga$_x$P$_{1-x}$As \cite{Marzari:1994} or Zn$_{x}$Mg$_{1-x}$S$_y$Se$_{1-y}$ \cite{Saitta:1998}, can be handled by straightforward extensions of this scheme.

I consider pseudopotential alchemy one of the feats of my early career, probably on par with DFPT. Unfortunately, its impact has been more modest: I did not have the vision to fully pursue its potential and, by the mid-nineties, I became increasingly drawn into the development of DFPT. Raffaele was on the verge of discovering the modern theory of polarization, which would make him famous and consume most of his energies for the rest of his professional life. As for Alfonso, I felt a nagging doubt that he did not value our work as much as I did, likely due to his proverbial understatement, as noted in Raffaele's contribution to this volume. I partially revised this view when he wrote a fine and widely cited review paper with Maria and Nadia \cite{Peressi:1998}, covering much of our work on the band offset problem. However, the truth is that our fine work, along with my own research on solid solutions-—-directly stemming from it—--has had only a limited impact. Yet, it remains very dear to me, and I still consider it a pillar of my own development as a scientist and as a man.

\bibliography{Alfonso}

\end{document}